\documentclass[twocolumn,floatfix,superscriptaddress,amsmath,showpacs,showkeys,aps,pre]{revtex4}

\usepackage[final]{graphicx}
\usepackage{acronym}
\usepackage{amsmath}
\usepackage{amssymb}

\def\msp{m_\text{sp}}
\def\Tsp{T_\text{sp}}
\def\taunuc{\tau_\text{nuc}}
\def\Cthr{C_\text{thr}}

\begin{document}

\title{Non-equilibrium Characterization of Spinodal Points using Short
  Time Dynamics}

\author{Ernesto S. Loscar}

\affiliation{Instituto de Investigaciones Fisicoqu\'\i{}micas
  Te\'oricas y Aplicadas (INIFTA) and Departamento de F\'\i{}sica,
  Facultad de Ciencias Exactas, Universidad Nacional de La Plata,
  c.c.~16, suc.~4, 1900 La Plata (Argentina)}

\affiliation{CCT La Plata, Consejo Nacional de Investigaciones
  Cient\'\i{}ficas y T\'ecnicas (CONICET, Argentina)}

\author{Ezequiel E. Ferrero}

\affiliation{Instituto de F\'{\i}sica de la Facultad de Matem\'atica,
  Astronom\'{\i}a y F\'{\i}sica (IFFAMAF-CONICET), Universidad
  Nacional de C\'ordoba, Ciudad Universitaria, 5000 C\'ordoba,
  Argentina}

\author{Tom\'as S. Grigera}

\affiliation{Instituto de Investigaciones Fisicoqu\'\i{}micas
  Te\'oricas y Aplicadas (INIFTA) and Departamento de F\'\i{}sica,
  Facultad de Ciencias Exactas, Universidad Nacional de La Plata,
  c.c.~16, suc.~4, 1900 La Plata (Argentina)}

\affiliation{CCT La Plata, Consejo Nacional de Investigaciones
  Cient\'\i{}ficas y T\'ecnicas (CONICET, Argentina)}

\author{Sergio A. Cannas}

\affiliation{Instituto de F\'{\i}sica de la Facultad de Matem\'atica,
  Astronom\'{\i}a y F\'{\i}sica (IFFAMAF-CONICET), Universidad
  Nacional de C\'ordoba, Ciudad Universitaria, 5000 C\'ordoba,
  Argentina}

\date{\today}

\begin{abstract}

  Though intuitively appealing, the concept of spinodal is rigorously
  defined only in systems with infinite range interactions (mean field
  systems). In short-range systems, a pseudo-spinodal can be defined
  by extrapolation of metastable measurements, but the point itself is
  not reachable because it lies beyond the metastability limit.  In
  this work we show that a sensible definition of spinodal points can
  be obtained through the short time dynamical behavior of the system
  deep inside the metastable phase, by looking for a point where the
  system shows critical behavior. We show that spinodal points
  obtained by this method agree both with the thermodynamical spinodal
  point in mean field systems and with the pseudo-spinodal point
  obtained by extrapolation of meta-equilibrium behavior in short
  range systems. With this definition, a practical determination can
  be achieved without regard for equilibration issues.

\end{abstract}



\maketitle

\acrodef{STD}{short-time dynamics}

\section{Introduction}
\label{Introduction}

First-order phase transitions are accompanied by hysteresis and
metastability: even though the thermodynamic transition happens at the
value $\varphi_t$ of the control variable $\varphi$, when $\varphi$ is
varied smoothly from $\varphi_i>\varphi_t$ to $\varphi_f<\varphi_t$,
the system remains in the phase corresponding to thermodynamic
equilibrium at $\varphi>\varphi_t$ (and conversely when changing
$\varphi$ in the opposite sense). When the phase survives being
carried out beyond its thermodynamic ``homeland'', it is called
\emph{metastable}. Metastable phases have a finite lifetime, but this
time can be very long. Diamond at room temperature and pressure, and
glass-forming supercooled liquids are very well-known examples of
long-lived metastable phases (so long-lived, in fact, that for many
purposes they can be considered an equilibrium phase). In general,
however, the metastable phase cannot exist for all $\varphi$, and it
is not observed if $\varphi_f$ is less than some value
$\varphi_\text{sp}$. This is the idea behind the concept of
\emph{spinodal point.}  However, to define the spinodal, some care is
required.

At the mean-field level, the spinodal is well defined. Focusing on the
ferromagnetic case to be specific, let's consider the extended free
energy per particle \cite{book:binney92} $f_3(T,m,h)$, with $h$ the
magnetic field and $m$ the magnetization. $f_3$ depends on two
conjugate variables because it is defined such that the probability of
finding a value $M$ of the magnetic moment is $\propto \exp[-\beta N
f_3(T,M/N,h)]$, where $N$ is the system size. In mean field, and in the
limit $N\to\infty$, $f_3$ has two minima as a function of $m$ for $T$
below some critical temperature $T_c$ and $h$ within some range
$-h_\text{sp}< h<h_\text{sp}$ \cite{review:binder87, book:binney92}.
As a function of $h$, the (first-order) transition occurs at
$h=h_c=0$.  When $h=0$, the two minima are symmetric, corresponding to
the two broken-symmetry phases.  When $h\neq0$, the absolute minimum
corresponds to the thermodynamic equilibrium (or stable) phase, while
the secondary local minimum defines a phase which is dynamically
stable but of higher free energy: the metastable phase.  When it
exists, the (mean field) metastable phase has infinite lifetime. At
$h=h_\text{sp}$, the secondary minimum disappears (it becomes an
inflection point) and the phase becomes unstable: a system prepared
with a magnetization different from the (thermodynamic) equilibrium
value evolves toward the equilibrium state. For $\lvert h
\rvert\ge h_\text{sp}$, $f_3$ has only one minimum.  Thus in mean field
the spinodal, which is the point where the metastable phase becomes
unstable (in the sense that a susceptibility becomes negative), is
also the limit of metastability, i.e.\ the point up to which the
metastable phase can be observed.

When the interactions have a finite range, matters are more
complicated \cite{review:binder87, spinodal:binder07}. On one hand,
the metastable phase ceases to be observable before it becomes
unstable \cite{review:kauzmann49,
  phase-transition-theory:patashinskii79,
  phase-transition-theory:patashinskii80}. This is because as the
system moves away from the transition, the lifetime of the metastable
phase decreases while its relaxation time increases. When they become
of the same order, the phase is unobservable. This is the
metastability limit, which is thus different from the spinodal. The
metastability limit is also called \emph{kinetic spinodal}
\cite{phase-transition-theory:kiselev01}, and the term
\emph{thermodynamic spinodal} is sometimes used for the spinodal as
defined above (onset of an instability). On the other hand, the order
parameter can fluctuate in space. Although $f_3$ can still be defined
(and can be used to compute the true equilibrium properties), in the
thermodynamic limit it has no convexity changes and only one minimum;
therefore, it is useless to define a spinodal. Further, although the
extensive $F_3$ has a double-well shape
\cite{metastability:larralde06}, the local maximum cannot be
interpreted as a barrier to the growth of the stable phase, because
the global magnetization is not a good coordinate to describe this
process: a supercritical nucleus of the stable phase (one whose growth
is thermodynamically favored) \cite{book:chaikin00} can form without
change in global magnetization \cite{metastability:larralde06}. The
eventual disappearance of the secondary minimum in $F_3$ is hence
unrelated to the spinodal.

The spinodal is then beyond the metastability limit, and hence outside
the realm of thermodynamics and of quasi-equilibrium treatments. Due
to these difficulties, it has been concluded that the spinodal only
makes sense in mean field \cite{spinodal:herrmann82} or in finite size
systems \cite{spinodal:binder07}. However, signs of an instability are
detectable in (meta)equilibrium measurements: the susceptibility and
relaxation times of the metastable phase increase as one goes deeper
into the metastable region, and if extrapolated with a power law, seem
to diverge at a point beyond the metastability limit, called
\emph{pseudospinodal} \cite{spinodal:herrmann82, potts:fernandez92}.

Here, rather than to abandon the idea of a spinodal in finite-range
systems, we propose to \emph{define} it through out-of-equilibrium
properties. The idea is that the spinodal should be a point with
infinite susceptibility and infinite relaxation time. In this sense it
resembles the critical point of a second-order phase transition. It is
known \cite{short-time:janssen89,review:zheng06} that it is possible
to detect critical points by studying the short-time dynamic behavior
of the order parameter and correlation functions.  We propose to use
the same method to identify a point deep into the metastable region
which, at short times, behaves dynamically like a critical point. This
pseudocritical regime (in the sense that it lasts only for a finite
time) does not imply the existence of a thermodynamic singularity (see
discussion in sec.~\ref{conclu}).  We show that in a mean-field model,
where the spinodal is well-defined and can be worked out analytically,
the point identified with this technique is precisely the
thermodynamic spinodal. Thus by defining the spinodal as the point
where this pseudocritical dynamics takes place we provide a sensible
generalization of the spinodal concept to finite-range systems.  We
apply the method to the two-dimensional $q$-state Potts model
\cite{review:wu82} with $q>4$, where it gives a reasonable result,
providing a bound for the metastability limit and locating the
spinodal very near to the pseudospinodal. The technique has the
advantage that it does not need equilibrium data, which is an
essential requirement in order to determine the pseudospinodal.

Our proposal was inspired by the finding of Sch\"ulke and Zheng
\cite{short-time:schulke00} that the \ac{STD} applied to the Potts
model defines two ``pseudocritical points'', which are closer together
the weaker the (first-order) transition, and coincide for second-order
transitions. A similar situation was observed in models with
out-of-equilibrium transitions \cite{short-time:monetti01,
  short-time:saracco03}.

The paper is organized as follows. In Sec.~\ref{STD} we briefly review
the \ac{STD} technique. In Sec.~\ref{meanfield} we consider mean-field
spin models and we show that the \ac{STD} method accurately describes
the thermodynamic spinodal in those models. In Sec.~\ref{Potts model}
we apply the method to the ferromagnetic $q$-state Potts model with
nearest-neighbor interactions in two dimensions. Sec.~\ref{conclu}
presents our conclusions.

\section{Using short-time dynamics to identify critical points}
\label{STD}

The STD technique to identify critical points has been reviewed in
Refs.\ [\onlinecite{review:zheng98, review:zheng06}]. Briefly, it is
based on the work of Janssen et al.\ \cite{short-time:janssen89}, who
studied model A (a $\phi^4$ Hamiltonian with Langevin dynamics) in the
out of equilibrium regime where correlation functions are still
nontrivial functions of two times and the order parameter is still
time-dependent. For the present work, the relevant result is that the
$n$-th moment of the order parameter
$m^{(n)}(t) = \,\bigl\langle \big{[} m(t)-\langle m(t) \rangle \big{]}^n \bigr\rangle $
obeys the scaling form\cite{short-time:janssen89, review:zheng98}
\begin{equation}
  m^{(n)}(t,\tau,L,m_0) = b^{-n\beta/\nu}
  g_n(b^{-z}t,b^{1/\nu}\tau, L/b, b^\mu m_0),
  \label{eq:short-time-general-scaling}
\end{equation}
where $t$ is time, $\tau$ is the reduced temperature
$\tau=(T-T_c)/T_c$, $L$ is the system size, $m_0$ is the initial value
of the order parameter (assumed nonzero but small), and $b$ is a
rescaling parameter. $\mu$ is a new universal exponent that describes
the short time behavior, while $\beta$, $\nu$, and $z$ are the usual
critical exponents \cite{book:binney92}.

From Eq.~(\ref{eq:short-time-general-scaling}), setting $b=t^{1/z}$,
for large values of $L$ and small values of $t^{1/z}m_0$ one obtains
\begin{equation}
  \label{eq:short-time-m-scaling}
  m(t,\tau,m_0) \sim m_0\, t^\theta F(t^{1/\nu z}\tau), \qquad \theta =
  \frac{\mu-\beta\nu} {z},
\end{equation}
so that precisely at the critical point $\tau=0$, the order parameter
obeys a power law $m(t)\sim t^\theta$. Similarly, setting $m_0=0$, one
gets for the second moment at the critical point
\begin{equation}
  \label{eq:scaling-second-moment}
  m^{(2)}(t) \sim t^{d/z-2\beta/z\nu}.
\end{equation}

It is generally assumed (and in agreement with numerical results
\cite{review:zheng98, review:zheng06}) that when the initial condition
is the ordered state ($m_0=1$), a scaling similar to
Eq.~(\ref{eq:short-time-general-scaling}) holds:
\begin{equation}
  m^{(n)}(t,\tau,L) = b^{-n\beta/\nu}
  g_n(b^{-z}t,b^{1/\nu}\tau, L/b).
  \label{eq:short-time-ordered-scaling}
\end{equation}
From this equation one obtains, for $m_0=1$ and large $L$,
\begin{equation}
  m(t) = t^{-\beta/\nu z} G(t^{1/\nu z}\tau),
  \label{eq:short-time-m-ordered}
\end{equation}
and taking the derivative of $\log m$,
\begin{equation}
  \left.\frac{\partial \log m(t,\tau)}{\partial \tau}\right|_{\tau=0}
  \sim t^{1/\nu z}.
  \label{eq:scaling-log-m}
\end{equation}

The critical point can then be obtained by performing Monte Carlo
simulations at several temperatures and looking for the value of $T$
at which the power laws in time predicted by
Eqs.~(\ref{eq:short-time-m-scaling}), (\ref{eq:scaling-second-moment})
and (\ref{eq:short-time-m-ordered}) hold. In addition, these equations
together with Eq.~(\ref{eq:scaling-log-m}) allow to determine the
critical exponents~\cite{review:zheng06}.

Here we apply the above method to look for singular behavior in the
metastable region of a first order phase transition. By tuning the
appropriate control parameter (external field or temperature) we look
for a value where the power laws (\ref{eq:short-time-m-scaling}),
(\ref{eq:scaling-second-moment}) and (\ref{eq:short-time-m-ordered})
hold at short time scales (for very long times, the approach to the
proper equilibrium phase is seen). This value of the control parameter
can be sensibly defined as a spinodal, as we show in the cases studied
below.

\section{Spinodal points in systems with long range interactions}
\label{meanfield}
In this section we consider spin models where each spin interacts with
every other spin. For those systems mean field theory is exact and
therefore provides a first check for the \ac{STD} method.  Let us
first briefly review the analytical mean-field results on the spinodal
points.

\subsection{Thermodynamic spinodals}
\paragraph{Curie-Weiss-Ising model.}
\label{The Curie-Weiss Ising model in a magnetic field}

We first consider the Curie-Weiss, or fully-connected, version of the
Ising model. In the presence of an external magnetic field $h$ the
Hamiltonian is given by
\begin{equation}
  {\cal H}_\text{CWI}=-\dfrac{J}{2N}\displaystyle\sum_{i\neq j} s_is_j
  -h\displaystyle\sum_{i=1}^{N}s_i ,
\end{equation}
where $N$ is the number of spins ($s_i=\pm1$), $h$ is an external
magnetic field and $J>0$. The extended free energy per spin, $f_{3}$,
can be computed exactly because ${\cal H}_{CW}$ is an explicit
function of the total magnetization $M=\sum_{i} s_i$, and the
partition function can be evaluated for fixed $M$. The result, in the
limit $N\to\infty$, is
\begin{align}
  f_3(T,m,h) = &
  \dfrac{1}{\beta} \biggl[ \frac{1+m}{2} \ln \frac{1+m}{2}
  +\frac{1-m}{2} \ln \frac{1-m}{2} \biggl] \nonumber \\
 & -\dfrac{J}{2}m^2-hm,
\end{align}
where $\beta=1/T$ (we take Boltzmann's constant equal to 1) and
$m=M/N$. The absolute minimum of $f_3$ with respect to $m$ defines the
stable (equilibrium) solution $m(T,h)$. The model shows a second order
transition at $h=h_c=0$ and critical temperature $T_c =J$.  For
$T<T_c$, there is a line of first order transitions at $h=0$, where
$m(T,h)$ is singular. However, an analytic continuation $m_+(T,h)$
from positive to negative $h$ exists as long as $|h|$ is not too big
(and conversely a continuation $m_-(T,h)$ from negative to positive
$h$). These continuations are the metastable states and correspond to
local minima of $f_3$. Thus the conditions
\begin{equation}\label{spinodal-cond}
  \frac{\partial f_3(T,m,h)}{\partial m}=0, \qquad
  \frac{\partial^2 f_3(T,m,h)}{\partial m^2}>0,
\end{equation}
define the (meta)stable states. The secondary minimum (and thus the
metastable solution) ceases to exist when $h < h_{\text{sp}}^{(-)}=-h_{\text{sp}}$, the
spinodal field, given by
\begin{equation}\label{spinodal-condhsp}
\left. \frac{\partial^2 f_3(T,m,h)}{\partial m^2}\right|_{h=h_{\text{sp}}^{(-)}} = 0.
\end{equation}

Since $\partial^2 f_3/\partial m^2=\chi_T^{-1}$, the susceptibility
diverges at $h_{\text{sp}}^{(-)}$, and it is straightforward to show that for
$h-h_{\text{sp}}^{(-)}\ll~1$
\begin{equation}
\chi_T \sim (h-h_{\text{sp}}^{(-)})^{-1/2}.
\end{equation}
In fact, this singularity can be treated like a usual critical point.
For example, for fixed $T$ we have
\begin{align}
C_h & \sim (h-h_{\text{sp}}^{(-)})^{-1/2} \\
\Delta m & \sim (h-h_{\text{sp}}^{(-)})^{1/2}
\end{align}
where $C_h$ is the specific heat and $\Delta m = m-m_{\text{sp}}^{(-)}$, with
$m_{\text{sp}}^{(-)}=m(h_{\text{sp}}^{(-)})$ (note that $\Delta m > 0$). Fixing $h$,
an expansion in $0<T_\text{sp}-T\ll1$ gives
\begin{align}
\Delta m & \sim (T_\text{sp}-T)^{1/2} \label{Deltam}\\
C_h & \sim (T_\text{sp}-T)^{-1/2} \\
\chi_T & \sim (T_\text{sp}-T)^{-1/2}\label{chi}
\end{align}

If we choose $\Delta m$ as order parameter, the singular behavior in
the neighborhood of the spinodal point can be characterized exactly as
in a true critical point. The set of critical exponents is
\begin{equation}
\beta=1/2,\qquad \alpha=1/2, \qquad \gamma=1/2, \qquad \delta=2,
\label{eq:exponents-mf-theory}
\end{equation}
which satisfy Rushbrooke and Widom scaling relations
\begin{align}
\alpha +2 \beta + \gamma & =2,\\
\gamma &=\beta ( \delta -1).
\end{align}
If we assume that $\eta=0$ like in mean field critical points, from
the Fisher scaling relation, $\gamma = \nu (2 - \eta)$,
we have $\nu=1/4$. Finally, from the Josephson hiperscaling,
\begin{equation}
\nu d =2 - \alpha,
\end{equation}
we can guess a critical dimension $d_c=6$, a result confirmed by a
renormalization group analysis \cite{spinodal:gunton78}.

\paragraph{Curie-Weiss-Potts model.}
The Curie-Weiss-Potts model is defined by the Hamiltonian
\begin{equation}\label{HPCW}
  {\cal H}_\text{CWP} = - \frac{J}{N} \sum_{i\neq j} \delta(\sigma_i,\sigma_j),
\end{equation}
where $J>0$, $\sigma_i=1,2,\ldots,q$ and $\delta(\sigma_i,\sigma_j)$
is the Kronecker delta. The extended free energy per spin $f_{3}$ for
this model can also be computed exactly \cite{review:wu82}
\begin{widetext}
\begin{equation}\label{fPCW-2}
  f_3(m,T) = - \frac{J(q-1)}{2\, q} m^2 +\frac{1}{\beta\,
    q}\left\{\left[ 1+ (q-1)\, m\right] \ln \left[ \frac{1+ (q-1)\,
        m}{q}\right] + (q-1)(1-m) \ln \left[ \frac{1- m}{q}\right]
  \right\}.
\end{equation}
\end{widetext}
The order parameter $m$ is defined as
\begin{equation}\label{mpotts}
 m = \frac{1}{q-1} \left(q
      \left<\max(m_k) \right>  -1 \right) ,
\end{equation}
where $m_k=\frac{1}{N}\displaystyle\sum_{i=1}^{N} \delta(\sigma_i,k)$,
$k=1\ldots q$, and the maximum means choosing the value of $k$ that
gives the highest value of $m_k$.  The extended free energy
(\ref{fPCW-2}) presents two local minima (and therefore a first order
transition) for $q \geq 3$, where the transition happens at $T_t= J\,
(q-2) \ln(q-1)/2(q-1)$ \cite{potts:mittag74}. A disordered metastable
solution with $m=0$ (supercooled paramagnet) exists for $\Tsp^{(-)} <
T < T_t$ (with $\Tsp^{(-)}=J/q$), as well as an ordered metastable
solution with $m=\msp \neq 0$ for $T_t < T < \Tsp^{(+)}$, where the
spinodal temperatures are obtained from Eqs.~(\ref{spinodal-condhsp})
with $h=0$.  An analysis similar to the above shows that close to the
upper spinodal $\Tsp^{(+)}$, the order parameter $\Delta m$, the
specific heat and the susceptibility show the same critical behavior
Eqs.~(\ref{Deltam})-(\ref{chi}) with the same critical
exponents~(\ref{eq:exponents-mf-theory}). Of course, the same set of
exponents is obtained from other mean field approximations: the Landau
$\phi ^4$ model and ---for the upper spinodal point--- the Landau
$\phi ^6$ model. The values of these exponents are confirmed by means
of renormalization group techniques \cite{spinodal:gunton78,
  spinodal:klein81}.

On the other hand, the behavior close to the lower spinodal
temperature $\Tsp^{(-)}$ is different. The susceptibility shows a
power-law divergence with exponent $\gamma=1$, while the magnetization
and the specific heat remain finite (in fact, they are identically
zero in the disordered phase). The vanishing of the specific heat is
peculiar to the approximation; the point is that in principle one
should not expect to find the same critical exponent in both
spinodals.

\subsection{Short time dynamics}
\label{STD-in-action}

We now apply the \ac{STD} procedure to the Curie-Weiss-Ising model in
the neighborhood of the spinodal point, using $\Delta m$ as an order
parameter. We consider the process at fixed temperature, with the
magnetic field as the control variable. We choose the initial
condition in the ordered state ($m=1$) corresponding to $h \to
\infty$. In this case, the procedure analog to a quench $T$ near
$T_c$ for a thermal second order transition is to set $h$ to a value
near $h=h_{\text{sp}}^{(-)}$. To determine $h_{\text{sp}}^{(-)}$, we assume a scaling like
Eq.~(\ref{eq:short-time-general-scaling}) with $\tau=(h-h_{\text{sp}}^{(-)})/h_{\text{sp}}^{(-)}$.

In this test case, we can check the spinodal field and magnetization
found with \ac{STD} against the exact values obtained from
Eqs.~(\ref{spinodal-cond}) and~(\ref{spinodal-condhsp})
\begin{align}
    m_{\text{sp}}^{(-)} &= \sqrt{1- \frac{T}{J}}, \label{msp} \\
    \beta h_{\text{sp}}^{(-)}  &= \frac{1}{2} \ln \left[
    \frac{1+m_{\text{sp}}^{(-)}}{1-m_{\text{sp}}^{(-)}} \right] - \frac{m_{\text{sp}}^{(-)} J}{T}.
    \label{hsp}
\end{align}

\begin{figure}
  \includegraphics*[scale=0.65]{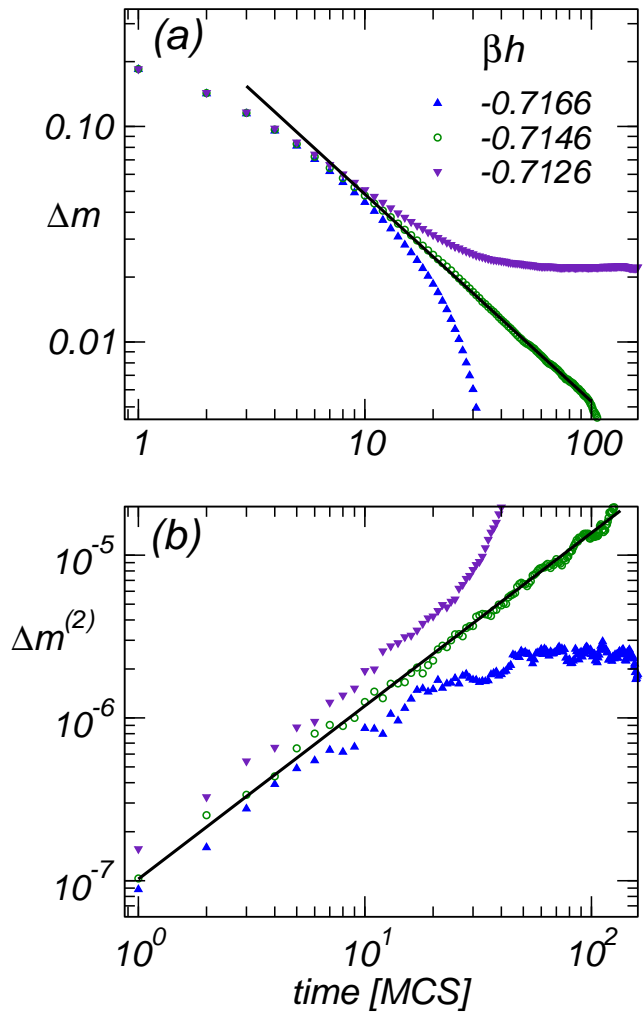}
  \caption{(Color online) Short time behavior of the first and second
    moment of $\Delta m$ for the Curie-Weiss-Ising model. Full
    (black) lines correspond to power law fits at $h=-0.7146=h_{\text{sp}}^{(-)}$.}
  \label{fig:fig1a}
\end{figure}

We have simulated the dynamics of this model using a standard
Metropolis algorithm in a system of $N=1.6 \times 10^6$ spins. We
started with all spins up and did $n\sim 10^3$ runs while recording
$\Delta m$ and the second moment of the magnetization per spin,
$\Delta m^{(2)}$.  Figure~\ref{fig:fig1a} shows the short time
behavior of $\Delta m$ and $\Delta m^{(2)}$ for $T=\frac{4}{9}T_c$,
where time is measured in Monte Carlo Steps (MCS, one MCS is
defined as a full cycle of $N$ spin update trials). From
Eqs.~(\ref{msp}) and~~(\ref{hsp}) we have $\beta h_{\text{sp}}^{(-)}\cong
-0.714627$ and $m_{\text{sp}}^{(-)}  \cong 0.745356$.

\begin{figure}
  \includegraphics*[scale=0.65]{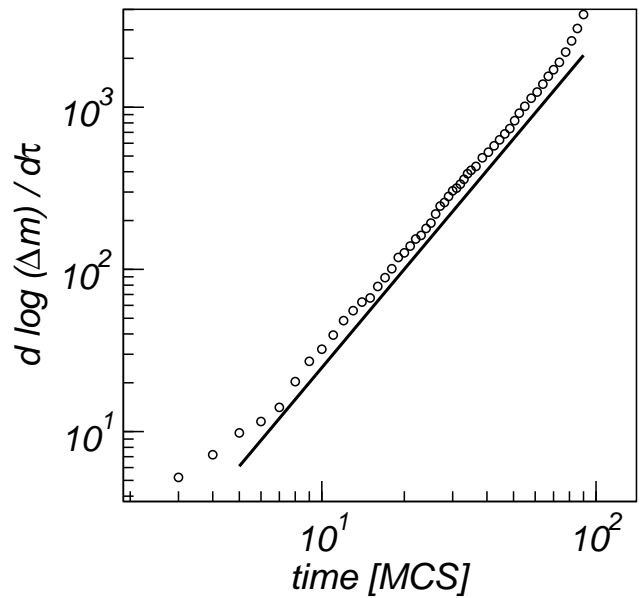}
  \caption{Short time behavior of the derivative of $\log \Delta m$
    with respect to the reduced field $\tau\equiv
    (h-h_{\text{sp}}^{(-)} / h_{\text{sp}}^{(-)} )$ evaluated at
    $\tau=0$; the straight line corresponds to a power law with
    exponent $w=2.02$.}
  \label{fig:figura2}
\end{figure}

For both the first and second moment
(Eqs.~(\ref{eq:scaling-second-moment})
and~~(\ref{eq:short-time-m-ordered})) a power law behavior $\Delta
m\sim t^{-x}$ and $\Delta m^{(2)}\sim t^y$ at $h=h_{\text{sp}}^{(-)}$
is found, with exponents $x=0.98\pm0.02$ and $y=1.03\pm0.02$
respectively.  The power laws are rather short-lived (lasting up to
$t\sim100$ MCS), and both observables deviate afterwards. However, as
we will show below this is a finite size effect, and a true power law
should be observerd in the thermodynamic limit.  In order to calculate
the derivative of $\log \Delta m$ from simulations we have taken the
magnetic field very close to the spinodal field, that is
$h=h_\text{sp} \pm \epsilon$ with $\epsilon=2\cdot 10^{-4}$.
Figure~\ref{fig:figura2} shows the numerical derivative obtained from
runs at three values of $h$.  Again these data can be fitted with a
power law with exponent $w =2.02\pm0.02$.

Using the scaling relations Eqs.~(\ref{eq:short-time-m-ordered})
and~(\ref{eq:scaling-log-m}) from sec.~\ref{STD} and the values of
$x$ and $w$, we obtain
\begin{equation}
\beta= 0.49 \pm 0.01,
\end{equation}
and assuming $d_c=6$, from Eqs.~(\ref{eq:scaling-second-moment})
and~(\ref{eq:short-time-m-ordered}) we have
\begin{equation}
z=2.01\pm0.03.
\end{equation}
The exponent $\nu$ is obtained from Eq.~(\ref{eq:scaling-log-m})
\begin{equation}
\nu=0.249\pm0.004.
\end{equation}
Thus, in this test case we find that the spinodal field as well as the
(static) critical exponents determined through \ac{STD} are in
excellent agreement with the theoretical results
(\ref{eq:exponents-mf-theory}), (\ref{msp}) and~(\ref{hsp}).

\begin{figure}
  \includegraphics*[scale=0.43]{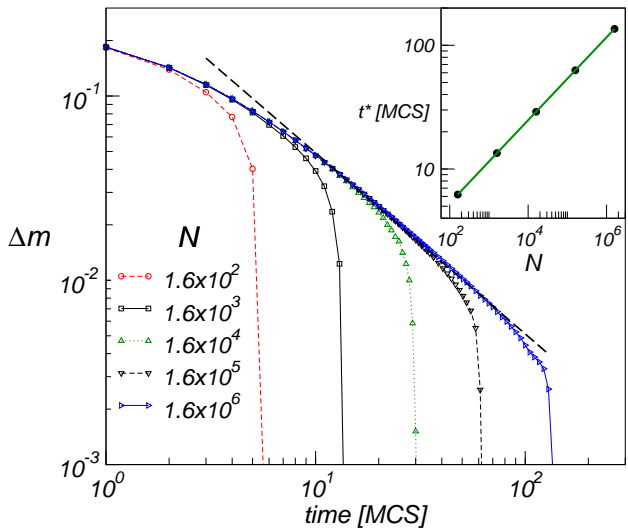}
  \caption{(Color online) Short time behavior of the order parameter
    $\Delta m$ at $h=h_{\text{sp}}^{(-)} $ for different system sizes
    $N$.  The dashed (black) line corresponds to a power law fit to
    the $N=1.6\times 10^6$ curve.  Inset: $t^*$, defined as the time
    for $\Delta m$ to reach 10$^{-3}$, vs.\ $N$.}
  \label{fig:figura3}
\end{figure}

We have also explored finite-size effects, which we use to obtain
another determination of $z$.  Figure~\ref{fig:figura3} shows the
evolution of $\Delta m$ for systems with number of spins ranging from
$N=1.6 \times 10^2$ to $N=1.6 \times 10^6$ at $h=h_{\text{sp}}^{(-)}$.
The same power law fit of Figure~\ref{fig:fig1a}a is also included
here.  There are clear finite size effects, and it is seen that the
power-law fit is valid for $t>t_\text{mic}\sim10$ up to a time which
increases with system size.  The time $t_\text{mic}$ is a microscopic
time scale and it is the time required sweep away the microscopic
short-wave behavior \cite{review:zheng98}. We define a time scale
$t^*$, so that for $t>t^*$ $\Delta m$ has abandoned the power law and
is rapidly evolving toward its equilibrium value (here we take $t^*$
as the time when $\Delta m = 3 \times 10^{-3}$).
Fig.~\ref{fig:figura3} (inset) shows $t^*$ vs.\ $N$ in a log-log
scale.  Such behavior can be understood by considering the evolution
of the correlation length $\xi$. In the critical regime, it increases
following the relation
\begin{equation}
\xi \propto t^{1/z}.
\end{equation}
We expect deviations from this critical law when the correlation
length becomes of the order of the linear dimension of the system,
$\xi \sim L$. If we assume that our system of $N$ spins behaves like
one with linear dimension $L= N^{1/d_c}$, the deviation would appear
when $\xi\sim N^{1/d_c}$, or $(t^*)^{1/z} \propto N^{1/d_c}$, and
therefore
\begin{equation}
t^* \propto N^u, \qquad u = \frac{z}{d_c}.
\end{equation}
In this way we can use finite size effects to extract $z$. From the
fit of Fig.~\ref{fig:figura3} (inset) we get $u=0.34\pm0.01$, which
together with $d_c=6$ gives $z=2.06\pm 0.06$, in excellent agreement
with the previous estimate.

\begin{figure}

\begin{center}
\includegraphics*[scale=0.45,angle=0]{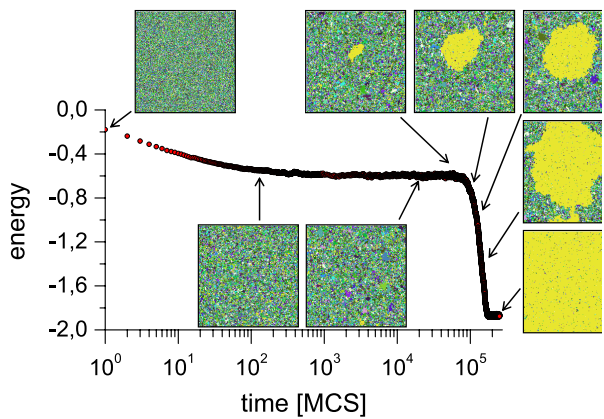}
\caption{\label{Efig2} (Color on-line) A typical single-sample energy
  per spin vs.\ time plot, after a quench from infinite temperature to
  $T=0.99 T_t$ for $q=24$ and $L=200$. Snapshots at selected times
  are also shown, with colors coding for spin values.}
\end{center}
\end{figure}

We conclude from this test that the \ac{STD} technique identifies the
(thermodynamic) spinodal points consistently with the static results.
We proceed now to a short-range model, where static approaches are
unsuitable.

\section{Spinodal points in the short-range Potts model}
\label{Potts model}

We consider now the nearest neighbor $q$-state Potts model
\cite{review:wu82} on the square lattice, with Hamiltonian
\begin{equation}
  {\cal H}_\text{P}=- J~\sum_{<i,j>}\delta\left(\sigma_i,\sigma_j\right), \qquad J>0,
  \label{Hamiltonian}
\end{equation}
where the sum runs now over all the pairs of nearest-neighbor sites.
The 2-$d$ Potts model undergoes a second order phase transition for
$q=2,3,4$ and a first order one for $q>4$. In the square lattice, the
transition temperature is known exactly to be\cite{potts:kihara54} $T_t(q)/ J
=1/\ln{\left(1+\sqrt{q}\right)}$. For $q$ larger
but near $q=4$, however, the transition is very weak. As Binder
\cite{potts:binder81} pointed out, the pseudospinodal temperatures
$\Tsp^{(+)}$ and $\Tsp^{(-)}$ are extremely close to $T_t$ ($[T_t -
\Tsp^{(-)}]/T_t \lesssim 10^{-3}$ for $q=5,6$). It is thus very hard
to establish whether the pseudospinodals are different from the
transition temperature in the thermodynamic limit (Binder
\cite{potts:binder81} concluded that systems at least as large as
$1000 \times 1000$ would be required), and the question of the
existence of metastable phases in the thermodynamic limit has remained
controversial \cite{potts:fernandez92, potts:meunier00}.

We have determined the lower spinodal of this model using the
short-time dynamics, as well as the pseudospinodal from
metaequilibrium measurements. We have done Monte Carlo simulations
with single-spin-flip Metropolis dynamics on square lattices with $N=L
\times L$ sites ($L$ ranging between $L=200$ and $L=4000$) and
periodic boundary conditions, for $q=9$, 12, 24, 48, 96 and 192. The
runs were quenches from infinite temperature, i.e.\ at constant $T$
(typically below $T_t(q)$) but starting from a random configuration.

\subsection{Pseudospinodal} \label{pseudospinodal}

\begin{figure}
\begin{center}
\includegraphics*[scale=0.35]{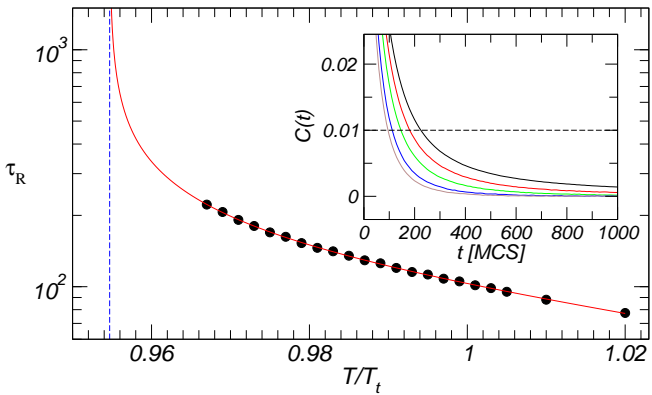}
\caption{(Color on-line) Relaxation time vs. $T/T_t$ for $L=1000$ and
  $q=96$. The continuous line is a fit to the data up to $T/T_t \leq 1.02$
  using the Eq.~(\ref{fittaur}). Error bars are smaller than the symbol 
  size. The inset shows the corresponding stationary correlation function
  vs. $t$ for temperatures ranging from $T=1.005\, T_t$ (left) to
  $T=0.967\, T_t$ (right).  The vertical dashed line is $T=T^*$.}
\label{correl}
\end{center}
\end{figure}

For a narrow range of quench temperatures $T<T_t$ the system gets
stuck in a high energy paramagnetic metastable state, where it remains
for some random time, after which it relaxes (relatively quickly) to
the equilibrium ferromagnetic state through a nucleation process. A
typical energy vs.\ time curve is shown in Fig.~\ref{Efig2}, together
with some snapshots of spin configurations, illustrating the
nucleation process.  The time for the formation of a critical nucleus
has a log-normal distribution, whose average $\taunuc$ depends both on~\cite{potts:ferrero07} 
$T$ and $L$. To find the pseudospinodal, we
look for a divergence of the relaxation time of the metaequilibrium
phase, so we must ensure that the measurements are done at times
shorter than $\taunuc(T)$ in order to avoid entering the regime where
domains of the stable phase have begun to grow. For this we consider
the two-time autocorrelation function
\begin{equation}
  C(t_1,t_2) = \frac{q}{q-1} \left\langle  \frac{1}{N} \sum_i^N
    \delta\left(\sigma_i(t_1),\sigma_i(t_2)\right) -
    \frac{1}{q} \right\rangle,
  \label{C(t,tw)}
\end{equation}
where $t_1$ is the time elapsed since the quench, $t_2>t_1$, and the
average is taken over different realizations of the thermal noise. Out
of equilibrium, $C(t_1,t_2)$ depends on both $t_1$ and $t_2$ while in
a stationary (meta)stable state it depends only on the time difference
$t\equiv t_2-t_1$. To compute the relaxation time $\tau_R$, we use
correlation data only from the regime where it is independent of
$t_1$, thus staying at temperatures above the metastability limit.
The typical behavior of $C(t)$ in this regime is shown in the inset of
Fig.~\ref{correl} for $q=96$. It is clear that the relaxation time is
growing as one goes deeper into the metastable region. We defined
$\tau_R$ as the time at which $C(t)$ falls below some threshold
$\Cthr$ (see inset of Fig.~\ref{correl}). Varying $\Cthr$ within a
reasonable range we obtained similar results; we have included this
arbitrariness in the error estimates.  The behavior of $\tau_R$ as a
function of temperature is plotted in Fig.~\ref{correl} for $q=96$. 

From these data we estimated the the pseudospinodal temperature
$T^*$. This is not an easy task: far from the divergence, corrections
to power-law scaling can be important, but on the other hand the
pseudospinodal cannot be approached with equilibrium measurements. The
estimates of $T^*$ and associated exponents will necessary have
relatively large uncertainties. To proceed, we fit the relaxation time
with a scaling form \cite{Goldenfeld},
\begin{equation}
  \tau_R=  A\left(\frac{T-T^*}{T_t}\right)^{-b} \left[1+B
    \left(\frac{T-T^*}{T_t}\right)^c \right],
  \label{fittaur}
\end{equation}
fixing $c$ to different values.  The best fit according to the
correlation coeficcient ($R^2=0.99989$) was obtained for $c=1$, for
which the scaling correction to the power law is around $5\%$. However
other values also give good fits.  We find that, while the estimates
of $b$ depend on the form of the correction chose, the estimate of
$T^*$ is within the interval $T^* = 0.95 \pm 0.01$. We cannot give an
accurate estimate of $b$ other than stating that the divergence is
weak, with $b$ ranging from $0.2$ to $0.85$. This is enough for our
purpose of comparing $T^*$ with the spinodal obtained from \ac{STD},
which we compute in the next section.

\subsection{The spinodal through \ac{STD}}

\begin{figure}
\begin{center}
\includegraphics*[scale=0.55]{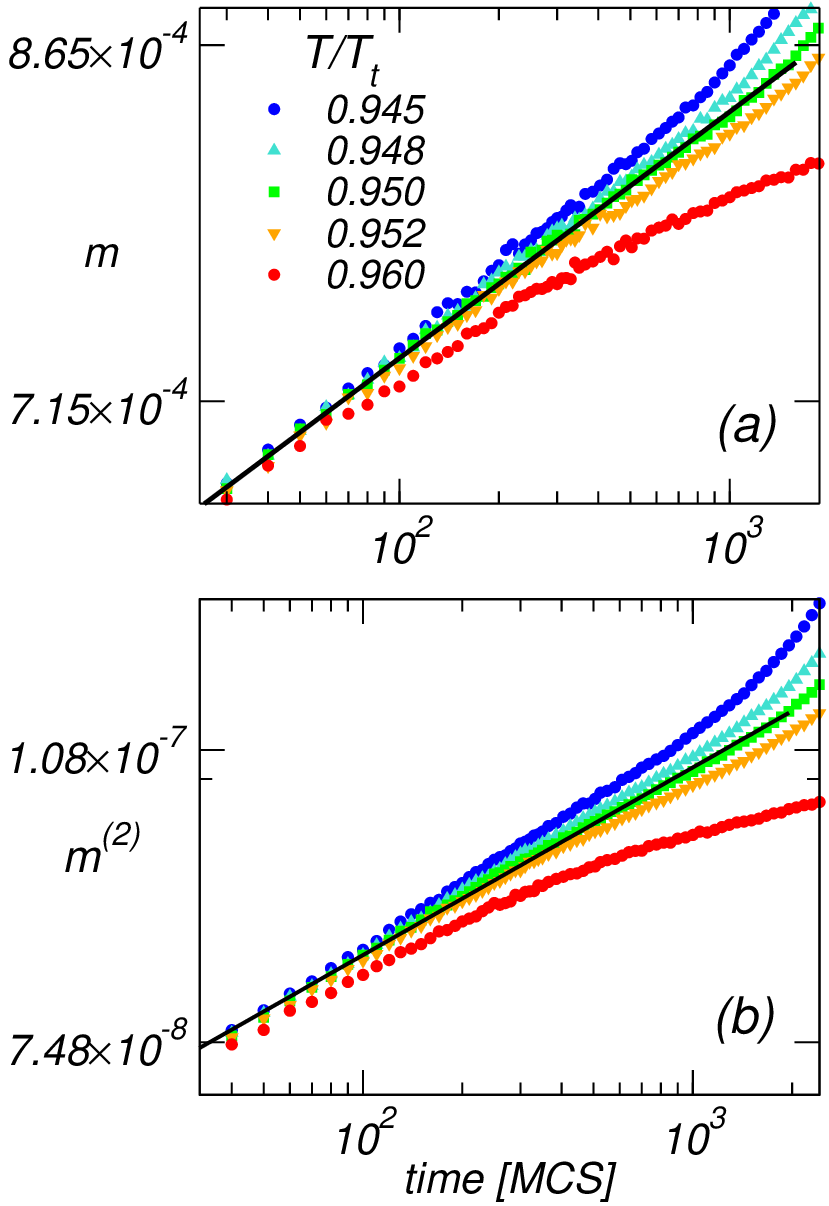}
\caption{(Color on-line) Short time behavior in a supercooled state at
  different temperatures for $q=96$ and $L=480$. Full lines are power
  law fittings for $T=0.950 T_t$. (a) Order parameter $m$; (b) second
  moment $m^{(2)}$. }
  \label{mmpotts}
\end{center}
\end{figure}

We now attempt to find the spinodal using the \ac{STD}. We consider
the dynamic behavior of the magnetization Eq.~(\ref{mpotts}) and the
corresponding second moment \cite{review:zheng98},
\begin{equation}
  m^{(2)} = \frac{q}{(q-1)^2}\sum_{j=1}^q \left\langle
    \left(\frac{1}{N}\sum_{i=1}^N\delta(\sigma_i,\sigma_j)  - \frac{1}{q}
    \right)^2\right\rangle,
\end{equation}
starting from a completely disordered state. The results for $q=96$
are shown in Fig.~\ref{mmpotts}. We observe a clear power law
increase, $m \sim t^\theta$ and $m^{(2)} \sim t^\omega$, spanning two
decades in time for $T=\Tsp^{(-)}=(0.950\pm 0.002)T_t$. The exponents
we find are rather small ($\theta\approx 0.06$ and $\omega\approx
0.1$).

\begin{figure}
\begin{center}
\includegraphics*[scale=0.33]{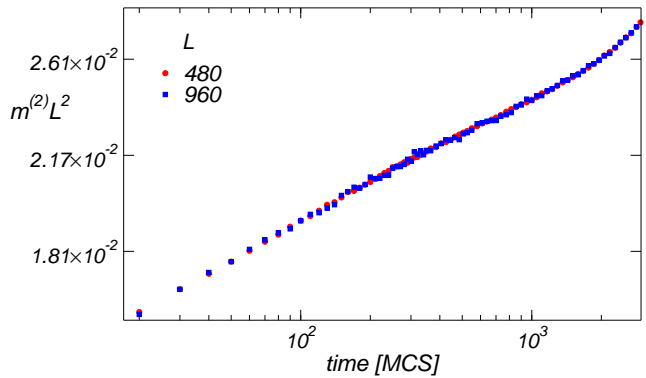}
\caption{(Color on-line) Short time behavior of the second moment
  $m^{(2)}$ at $T=0.95\, T_t$ for different system sizes ($L$) and
  $q=96$ Potts model.}
  \label{mm2pottsL}
\end{center}
\end{figure}

At variance with the mean-field case, the \ac{STD} results show no
noticeable finite-size effects. Fig.~\ref{mm2pottsL} shows $m^{(2)}
L^2$ vs.\ time at the spinodal temperature $\Tsp^{(-)}=0.950 T_t$ for
$L=480$ and $960$. The curves are almost indistinguishable, indicating
that the spinodal critical regime lasts for a size-independent time.
Clearly, at this time correlations have not yet reached lengths of the
order of the smallest system size. This is in contrast to the
mean-field case (Fig.~\ref{fig:figura3}), where it seems that the
number of spins that must become correlated before faster growth of
the stable phase begins grows with $N$.  Again, we see that the span
of the power law is limited, so that the values of the exponents are
not very accurate. However, our interest is to establish the existence
of a power-law-like regime to define $\Tsp$ and to compare its value
with the pseudospinodal.

\begin{figure}
\begin{center}
\includegraphics*[scale=0.33]{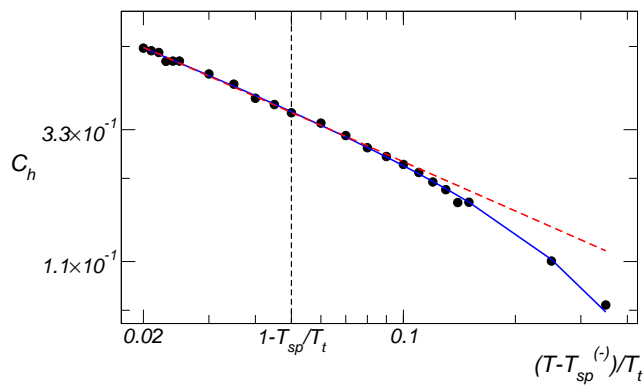}
\caption{(Color on-line) Specific heat as function of $T-T_{sp}^{(-)}$
  in the metastable state for $q=96$ and $L=480$. The full (blue) line
  is a fit with a law $A\left(T-\Tsp^{(-)}\right)^\alpha
  \left[1+B\left(T-\Tsp^{(-)}\right)\right]$.  The dashed (red) line
  is a power law fit using data for $\left(T-\Tsp^{(-)}\right)/T_t <
  0.15$. Error bars are smaller than the symbol size. The vertical
  dashed line is $T=T_t$.  }
  \label{specific-heat-potts}
\end{center}
\end{figure}

\begin{figure}
\begin{center}
\includegraphics*[scale=0.33]{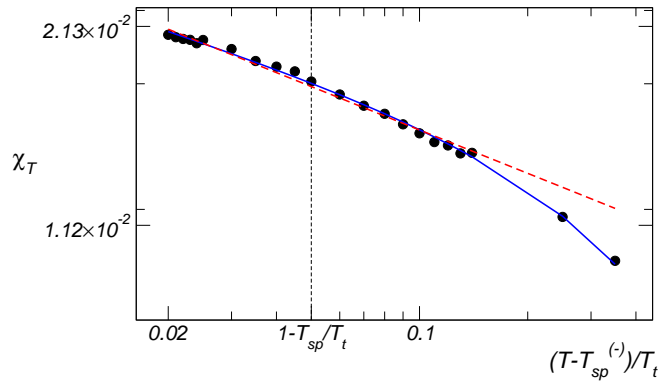}
\caption{(Color on-line) Magnetic susceptibility as function of
  $T-T_{sp}^{(-)}$ in the metastable state for $q=96$ and $L=480$. The
  full (blue) line is a fit with a law
  $A\left(T-\Tsp^{(-)}\right)^\alpha
  \left[1+B\left(T-\Tsp^{(-)}\right)\right]$.  The dashed (red) line
  is a power law fit using data for $\left(T-\Tsp^{(-)}\right)/T_t <
  0.15$. Error bars are smaller than the symbol size.  The vertical
  dashed line is $T=T_t$.  }
  \label{susceptibility-potts}
\end{center}
\end{figure}

The temperature $\Tsp^{(-)}$ identified by \ac{STD} is equal to the
pseudospinodal (here defined as the apparent divergence of the
relaxation time). The near-instability of the system at $\Tsp^{(-)}$
also shows up in the specific heat $C_h $ and magnetic susceptibility
$\chi_T$. We computed these quantities in the metastable regime, where
we find the same values either through fluctuations the energy
(magnetization) or by a numerical derivative with respect to
temperature (field). Plots for $q=96$ are shown in
Figs.~\ref{specific-heat-potts} and~\ref{susceptibility-potts}.  $C_h$
and $\chi_T$ grow in a way compatible with a divergence at
$T=\Tsp^{-}$. However, the available equilibrium data span too short a
range to get a reliable estimate of the critical exponents.

Let us remark that that Fig.~\ref{mmpotts} is not completely
equivalent to Fig.~\ref{fig:fig1a} (Curie-Weiss Ising). The second
moment of the order parameter (part b of both figures) behaves the
same in both cases, but not the order parameter itself. This is
because of the initial conditions that must be used. In general, at a
critical point, if one starts with a large value of $\Delta m$, one
observes a power-law decay towards zero (the Ising spinodal
corresponds to this case). If one starts from near zero $\Delta m$,
the order parameter is first observed to \emph{increase} with a power
law, then to decay again to zero \cite{short-time:janssen89}. This
\emph{initial increase} is governed by an exponent $\theta$, which
cannot be determined from the equilibrium critical exponents alone
(see eq.~(\ref{eq:short-time-m-scaling})) \cite{short-time:janssen89,
  review:zheng98}. The latter is the situation in the Potts model at
$T=\Tsp^{(-)}$. In an equilibrium critical point, one is free to
choose the initial condition, so both situations can be observed. In
the Potts $\Tsp^{(-)}$ case, however, we cannot start with a high
value of $m$, because that would place us automatically outside the
metastable phase we are trying to observe, so we can only hope to see
the initial increase (the corresponding decrease at longer times is
masked by the evolution to the stable phase). In the Curie-Weiss Ising
case, we could in principle set $\Delta m \approx 0$ and see something
similar to Fig.~\ref{mmpotts}a. We have failed to observe such
increase, however, indicating that the exponent $\theta$ is
zero. Indeed, in equilibrium it is known that $\theta=0$ in the
mean-field case \cite{short-time:janssen89}.

\begin{center}
\begin{figure}
  \includegraphics*[scale=0.57]{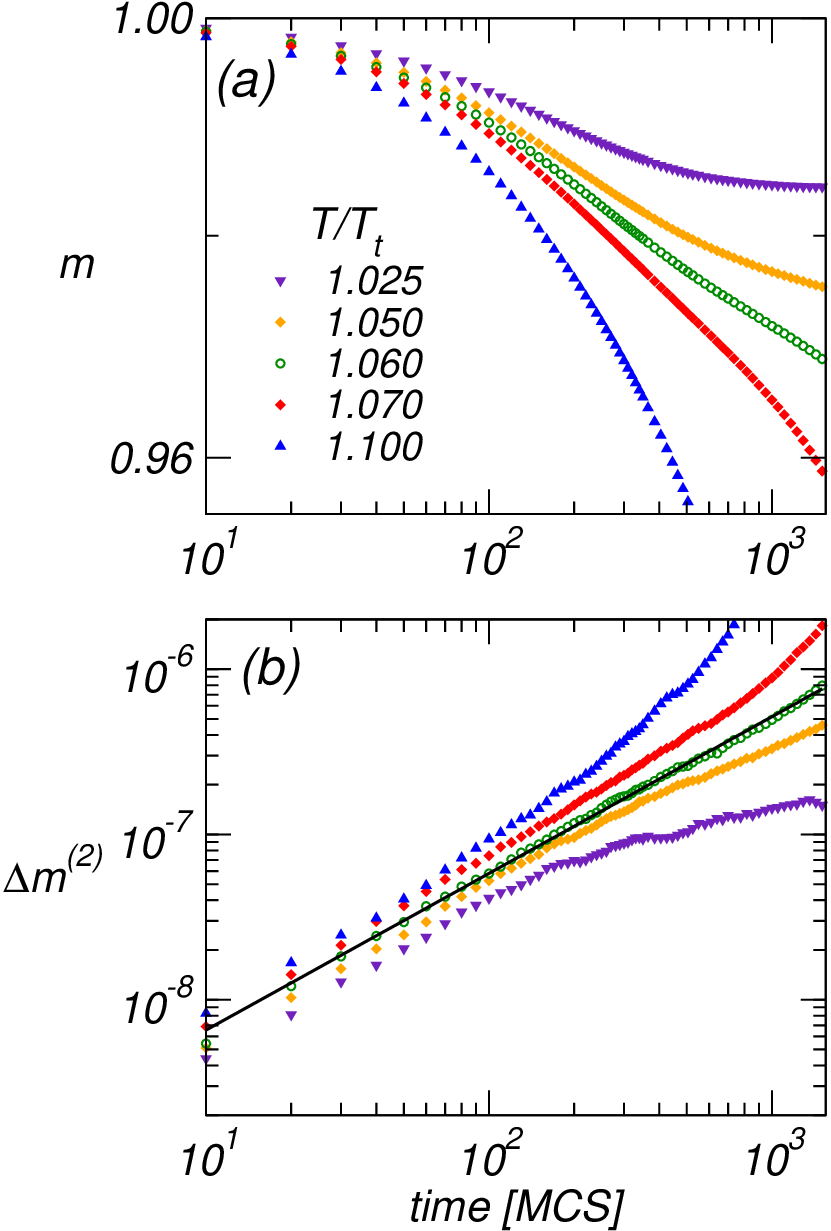}
  \caption{(Color on-line) Short time behavior in a superheated state
    at different temperatures for $q=96$ and $L=480$. Full line in (b)
    is a power-law fit for $T=1.060 T_t$. (a) Order parameter $m$; (b)
    second moment $m^{(2)}$.}
  \label{fig:Potts-upper}
\end{figure}
\end{center}

On the other hand, in the upper spinodal point of the the Potts model,
$T=\Tsp^{(+)} > T_t$, we can start with an ordered initial state, and
we should again observe $m$ \emph{decrease}. Indeed, we have simulated
the short time dynamics for $q=96$ and $N=480$ taking an ordered
initial state and tuning the temperature above the transition value
$T_t$. Fig.~\ref{fig:Potts-upper} shows the magnetization and its
second moment measured in the same way as in Fig.~\ref{fig:fig1a} for
different temperatures. In Fig.~\ref{fig:Potts-upper}b we can identify
a nice power law in the second moment for $\Tsp^{(+)}/T_r = 1.06 \pm
0.01$ over more than two decades. The fitted exponent is $y \approx 0.90$.
At this temperature, Fig.~\ref{fig:Potts-upper}a shows the
magnetization decreases in the same way as in Fig.~\ref{fig:fig1a}a.
The main difference is that the spinodal magnetization in the latter
case is known exactly, while in the Potts case it is unknown and
cannot be estimated with good precision with these data. Thus we do
not see a power-law in the $m$ plot, and we rely on $m^{(2)}$ to find
the spinodal.

\begin{figure}
\begin{center}
\includegraphics*[scale=0.36]{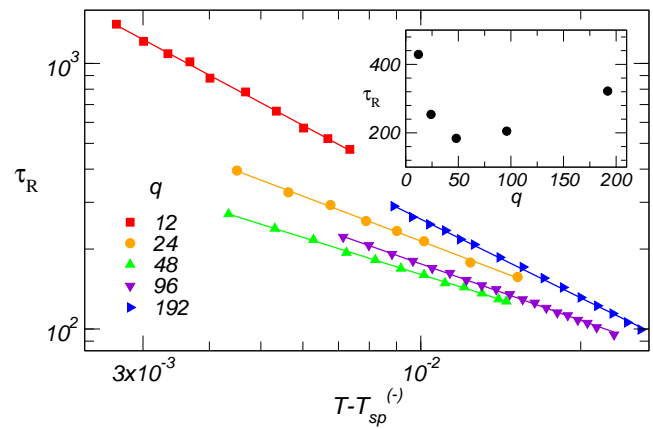}
\caption{(Color on-line) Relaxation time as function of $T-\Tsp^{(-)}$
  for $L=1000$ and different values of $q$. Error bars are smaller
  than symbol sizes. Note however that due to the error in the
  determination of $\Tsp$, the points could be uniformly shifted
  horizontally. The continuous lines are power law fits. Inset:
  $\tau_R$ as a function of $q$ for a fixed temperature
  $T-T_{sp}=0.008$.}
  \label{tau-q}
\end{center}
\end{figure}

Finally, we have repeated the above simulations for several values of
$q$, computing the relaxation time and the \ac{STD} behavior. In all
cases we found that the spinodal $\Tsp^{(-)}$ found by \ac{STD} is
compatible with the pseudospinodal $T^*$ from metaequilibrium
measurements. The relaxation time as a function of temperature for
several $q$ and $L=1000$ is given in Fig.~\ref{tau-q} (the same
results were found for $L=2000$ and $4000$). This calculation was done
using the same threshold $\Cthr=0.01$ for all values of $q$.  We
observed that for fixed $T-\Tsp$, the relaxation time is non-monotonic
as a function of $q$, with a minimum around $q\approx 50$ (see inset
of Fig.~\ref{tau-q}), and the same non-monotonic behavior is observed
in the exponents. We do not have an explanation for this
behavior. However, the growth of the relaxation time for large values
of $q$ is consistent with the appearance of a true singularity, since
the $q\rightarrow \infty$ limit is mean-field\cite{pearce80}. We also
observed that the kinetic spinodal is strongly $q$-dependent and
therefore the minimum distance to $T_{sp}^{(-)}$ available for
metaequilibrium measurements increases with $q$.

v
In Fig.~\ref{Tsp-potts} we show $(T_t-\Tsp^{(-)})/T_t$ vs.\ $q$ (we
have included the ``pseudocritical'' temperatures found in
Ref.~\onlinecite{short-time:schulke00} for $q=5,7$). We see that
$\Tsp^{(-)}$ systematically departs from $T_t$ as $q$ increases. In
fact, the data are very well fitted by the logarithmic form $A \log^a
(1+q-4)$ with $a=2.81$, in qualitative agreement with the behavior of
the mean field or Curie-Weiss-Potts, and the Bethe lattice solution
with coordination number~\cite{potts:peruggi83, potts:wagner00} $z=3$,
as expected in the large $q$ limit\cite{pearce80}.

\section{Conclusions}
\label{conclu}

We have shown that it is possible to define the spinodal point through
the short-time dynamic behavior. The \ac{STD} can be used to detect a
point in the phase diagram where the dynamics is critical (albeit for
a finite time). In mean-field systems this coincides with the
thermodynamic spinodal defined through the vanishing of second
derivatives of the free energy, while in finite-dimensional systems it
serves as a definition of spinodal, a point where (meta)equilibrium
measurements are impossible (since it is beyond the metastability
limit, or kinetic spinodal). In the 2-$d$ Potts model, we found that
the spinodal defined in this way coincides with the pseudospinodal
found through fitting and extrapolation of metaequilibrium relaxation
times. Our results are consistent with the scaling behavior associated
with a growing correlation length. In particular, (pseudo)critical
exponents can be measured using \ac{STD}, which in the mean-field case
we have checked with the analytical result.

For the Potts model, this method gives a spinodal temperature
different from the transition temperature at all $q$ where the
transition is first order, even in the thermodynamic limit. Since the
spinodal provides only a bound for the metastability limit, this does
not settle the question of the existence of a metastable phase, but it
does show that the apparent convergence of spinodal and transition
temperatures is due to the extremely weak nature of the transition for
low values of $q$.

Appart from the conceptual advantage of allowing a definition of
spinodal points avoiding equilibration issues, the \ac{STD} method may
prove to be practically useful in establishing bounds for
metastability limits. This would be especially welcome for instance in
systems with very slow dynamics such as glassforming liquids, where
metaequilibrium measurements are out of the question (a direction in
which some work is in progress). And even more so if this technique
could be implemented experimentally.

Finally, let us remark that although our results show that one can
define a spinodal (which we call thermodynamic to distinguish it from
the kinetic spinodal) through a critical-like dynamics, this behaviour
is transient (except in mean-field), so that we do not conclude that
any thermodynamic potential is singular at this point. The
thermodynamic spinodal in this sense is better understood as similar
to the pseudospinodal, but defined in a way free from the
equilibration and extrapolation issues that pervade the
pseudospinodal.  In the 2-$d$ Ising model, the pseudospinodal has been
shown to be related to zeros of the partition function at
\emph{complex} values of temperature and field \cite{Gulbahce2004},
which approach the real plane as the range of the interaction is
increased.  We may conjecture a similar scenario for the Potts model,
whose thermodynamical behavior can be determined by the zeros of its
partition function in the complex temperature plane\cite{Kenna1998},
and where evidence has been found of singular behavior (with divergent
thermodynamical quantities) associated to some of those zeros
\cite{Matveev1996} for $q>4$. It is tempting to conjecture that the
(pseudo)critical dynamics observed here is associated to such zeros,
and that its lifetime is longer the nearer these zeros are to the real
plane. This remains to be studied, however, but it is an interesting
issue to investigate especially in short range interacting systems,
where a pseudo-singularity is observed even if the mechanism of a
vanishing free energy barrier for long wave length fluctuations is in
principle excluded.

\acknowledgements

We thank Andrea Cavagna, Irene Giardina and V\'\i{}ctor
Mart\'\i{}n-Mayor for many discussions on the issue of metastability
and spinodals, and M. Ib\'a\~nez de Berganza for useful discussions
about the Potts model. This work was supported by CONICET, Universidad
Nacional de C\'ordoba, Universidad Nacional de La Plata, ANPCyT grants
PICT 20472/04, PICT 33305/05 (Argentina) and ICTP (Italy).


\begin{thebibliography}{28}
\expandafter\ifx\csname natexlab\endcsname\relax\def\natexlab#1{#1}\fi
\expandafter\ifx\csname bibnamefont\endcsname\relax
  \def\bibnamefont#1{#1}\fi
\expandafter\ifx\csname bibfnamefont\endcsname\relax
  \def\bibfnamefont#1{#1}\fi
\expandafter\ifx\csname citenamefont\endcsname\relax
  \def\citenamefont#1{#1}\fi
\expandafter\ifx\csname url\endcsname\relax
  \def\url#1{\texttt{#1}}\fi
\expandafter\ifx\csname urlprefix\endcsname\relax\def\urlprefix{URL }\fi
\providecommand{\bibinfo}[2]{#2}
\providecommand{\eprint}[2][]{\url{#2}}

\bibitem[{\citenamefont{Binney et~al.}(1992)\citenamefont{Binney, Dowrick,
  Fisher, and Newman}}]{book:binney92}
\bibinfo{author}{\bibfnamefont{J.~J.} \bibnamefont{Binney}},
  \bibinfo{author}{\bibfnamefont{N.~J.} \bibnamefont{Dowrick}},
  \bibinfo{author}{\bibfnamefont{A.~J.} \bibnamefont{Fisher}},
  \bibnamefont{and} \bibinfo{author}{\bibfnamefont{M.~E.~J.}
  \bibnamefont{Newman}}, \emph{\bibinfo{title}{The theory of critical
  phenomena}} (\bibinfo{publisher}{Oxford University press},
  \bibinfo{year}{1992}).

\bibitem[{\citenamefont{Binder}(1987)}]{review:binder87}
\bibinfo{author}{\bibfnamefont{K.}~\bibnamefont{Binder}},
  \bibinfo{journal}{Rep. Progr. Phys.} \textbf{\bibinfo{volume}{50}},
  \bibinfo{pages}{783} (\bibinfo{year}{1987}),
  \urlprefix\url{http://stacks.iop.org/0034-4885/50/783}.

\bibitem[{\citenamefont{Binder}(2007)}]{spinodal:binder07}
\bibinfo{author}{\bibfnamefont{K.}~\bibnamefont{Binder}},
  \bibinfo{journal}{Philos. Mag. Lett.} \textbf{\bibinfo{volume}{87}},
  \bibinfo{pages}{799} (\bibinfo{year}{2007}).

\bibitem[{\citenamefont{Kauzmann}(1949)}]{review:kauzmann49}
\bibinfo{author}{\bibfnamefont{W.}~\bibnamefont{Kauzmann}},
  \bibinfo{journal}{Chem. Rev.} \textbf{\bibinfo{volume}{43}},
  \bibinfo{pages}{219} (\bibinfo{year}{1949}).

\bibitem[{\citenamefont{Patashinskii and
  Shumilo}(1979)}]{phase-transition-theory:patashinskii79}
\bibinfo{author}{\bibfnamefont{A.~Z.} \bibnamefont{Patashinskii}}
  \bibnamefont{and} \bibinfo{author}{\bibfnamefont{B.~I.}
  \bibnamefont{Shumilo}}, \bibinfo{journal}{Sov. Phys. {JETP}}
  \textbf{\bibinfo{volume}{50}}, \bibinfo{pages}{712} (\bibinfo{year}{1979}).

\bibitem[{\citenamefont{Patashinskii and
  Shumilo}(1980)}]{phase-transition-theory:patashinskii80}
\bibinfo{author}{\bibfnamefont{A.~Z.} \bibnamefont{Patashinskii}}
  \bibnamefont{and} \bibinfo{author}{\bibfnamefont{B.~I.}
  \bibnamefont{Shumilo}}, \bibinfo{journal}{Sov. Phys. Solid State}
  \textbf{\bibinfo{volume}{22}}, \bibinfo{pages}{655} (\bibinfo{year}{1980}).

\bibitem[{\citenamefont{Kiselev and
  Ely}(2001)}]{phase-transition-theory:kiselev01}
\bibinfo{author}{\bibfnamefont{S.~B.} \bibnamefont{Kiselev}} \bibnamefont{and}
  \bibinfo{author}{\bibfnamefont{J.~F.} \bibnamefont{Ely}},
  \bibinfo{journal}{Physica A} \textbf{\bibinfo{volume}{299}},
  \bibinfo{pages}{357} (\bibinfo{year}{2001}).

\bibitem[{\citenamefont{Larralde et~al.}(2006)\citenamefont{Larralde, Leyvraz,
  and Sanders}}]{metastability:larralde06}
\bibinfo{author}{\bibfnamefont{H.}~\bibnamefont{Larralde}},
  \bibinfo{author}{\bibfnamefont{F.}~\bibnamefont{Leyvraz}}, \bibnamefont{and}
  \bibinfo{author}{\bibfnamefont{D.~P.} \bibnamefont{Sanders}},
  \bibinfo{journal}{J. Stat. Mech.} \textbf{\bibinfo{volume}{2006}},
  \bibinfo{pages}{P08013} (\bibinfo{year}{2006}).

\bibitem[{\citenamefont{Chaikin and Lubensky}(2000)}]{book:chaikin00}
\bibinfo{author}{\bibfnamefont{P.~M.} \bibnamefont{Chaikin}} \bibnamefont{and}
  \bibinfo{author}{\bibfnamefont{T.~C.} \bibnamefont{Lubensky}},
  \emph{\bibinfo{title}{Principles of condensed matter physics}}
  (\bibinfo{publisher}{Cambridge University Press}, \bibinfo{year}{2000}).

\bibitem[{\citenamefont{Herrmann et~al.}(1982)\citenamefont{Herrmann, Klein,
  and Stauffer}}]{spinodal:herrmann82}
\bibinfo{author}{\bibfnamefont{D.~W.} \bibnamefont{Herrmann}},
  \bibinfo{author}{\bibfnamefont{W.}~\bibnamefont{Klein}}, \bibnamefont{and}
  \bibinfo{author}{\bibfnamefont{D.}~\bibnamefont{Stauffer}},
  \bibinfo{journal}{Phys. Rev. Lett.} \textbf{\bibinfo{volume}{49}},
  \bibinfo{pages}{1262} (\bibinfo{year}{1982}).

\bibitem[{\citenamefont{Fern\'andez et~al.}(1992)\citenamefont{Fern\'andez,
  Ruiz-Lorenzo, Lombardo, and Taranc\'on}}]{potts:fernandez92}
\bibinfo{author}{\bibfnamefont{L.~A.} \bibnamefont{Fern\'andez}},
  \bibinfo{author}{\bibfnamefont{J.~J.} \bibnamefont{Ruiz-Lorenzo}},
  \bibinfo{author}{\bibfnamefont{M.~P.} \bibnamefont{Lombardo}},
  \bibnamefont{and}
  \bibinfo{author}{\bibfnamefont{A.}~\bibnamefont{Taranc\'on}},
  \bibinfo{journal}{Phys. Lett. B} \textbf{\bibinfo{volume}{277}},
  \bibinfo{pages}{485} (\bibinfo{year}{1992}).

\bibitem[{\citenamefont{Janssen et~al.}(1989)\citenamefont{Janssen, Schaub, and
  Schmittmann}}]{short-time:janssen89}
\bibinfo{author}{\bibfnamefont{H.~K.} \bibnamefont{Janssen}},
  \bibinfo{author}{\bibfnamefont{B.}~\bibnamefont{Schaub}}, \bibnamefont{and}
  \bibinfo{author}{\bibfnamefont{B.}~\bibnamefont{Schmittmann}},
  \bibinfo{journal}{Z. Phys. B} \textbf{\bibinfo{volume}{73}},
  \bibinfo{pages}{539} (\bibinfo{year}{1989}).

\bibitem[{\citenamefont{Zheng}(2006)}]{review:zheng06}
\bibinfo{author}{\bibfnamefont{B.}~\bibnamefont{Zheng}}, in
  \emph{\bibinfo{booktitle}{Computer simulation studies in condensed-matter
  physics}}, edited by \bibinfo{editor}{\bibfnamefont{D.~P.}
  \bibnamefont{Landau}}, \bibinfo{editor}{\bibfnamefont{S.~P.}
  \bibnamefont{Lewis}}, \bibnamefont{and} \bibinfo{editor}{\bibfnamefont{H.-B.}
  \bibnamefont{Sch\"uttler}} (\bibinfo{publisher}{Springer},
  \bibinfo{year}{2006}).

\bibitem[{\citenamefont{Wu}(1982)}]{review:wu82}
\bibinfo{author}{\bibfnamefont{F.~Y.} \bibnamefont{Wu}}, \bibinfo{journal}{Rev.
  Mod. Phys.} \textbf{\bibinfo{volume}{54}}, \bibinfo{pages}{235}
  (\bibinfo{year}{1982}).

\bibitem[{\citenamefont{Sch\"ulke and Zheng}(2000)}]{short-time:schulke00}
\bibinfo{author}{\bibfnamefont{L.}~\bibnamefont{Sch\"ulke}} \bibnamefont{and}
  \bibinfo{author}{\bibfnamefont{B.}~\bibnamefont{Zheng}},
  \bibinfo{journal}{Phys. Rev. E} \textbf{\bibinfo{volume}{62}},
  \bibinfo{pages}{7482} (\bibinfo{year}{2000}).

\bibitem[{\citenamefont{Monetti and Albano}(2001)}]{short-time:monetti01}
\bibinfo{author}{\bibfnamefont{R.~A.} \bibnamefont{Monetti}} \bibnamefont{and}
  \bibinfo{author}{\bibfnamefont{E.~V.} \bibnamefont{Albano}},
  \bibinfo{journal}{Europhys. Lett.} \textbf{\bibinfo{volume}{56}},
  \bibinfo{pages}{400} (\bibinfo{year}{2001}).

\bibitem[{\citenamefont{Saracco and Albano}(2003)}]{short-time:saracco03}
\bibinfo{author}{\bibfnamefont{G.~P.} \bibnamefont{Saracco}} \bibnamefont{and}
  \bibinfo{author}{\bibfnamefont{E.~V.} \bibnamefont{Albano}},
  \bibinfo{journal}{J. Chem. Phys.} \textbf{\bibinfo{volume}{118}},
  \bibinfo{pages}{4157} (\bibinfo{year}{2003}),
  \urlprefix\url{http://link.aip.org/link/?JCP/118/4157/1}.

\bibitem[{\citenamefont{Goldenfeld}(1992)\citenamefont{Goldenfeld}}]{Goldenfeld}
\bibinfo{author}{\bibfnamefont{N.} \bibnamefont{Goldenfeld}},
 \emph{\bibinfo{title}{Lectures on Phase Transitions and the Renormalization Group}} (\bibinfo{publisher}{Addison--Wesley},
  \bibinfo{year}{1992}).


\bibitem[{\citenamefont{Matveev and Shrock}(1996)}]{Matveev1996}
\bibinfo{author}{\bibfnamefont{V.} \bibnamefont{Matveev}} \bibnamefont{and}
  \bibinfo{author}{\bibfnamefont{R.} \bibnamefont{Shrock}},
  \bibinfo{journal}{Phys. Rev. E} \textbf{\bibinfo{volume}{54}},
  \bibinfo{pages}{6174} (\bibinfo{year}{1996}).


\bibitem[{\citenamefont{Kenna}(1998)}]{Kenna1998}
  \bibinfo{author}{\bibfnamefont{R.} \bibnamefont{Kenna}},
  \bibinfo{journal}{J. Phys. A} \textbf{\bibinfo{volume}{31}},
  \bibinfo{pages}{9419} (\bibinfo{year}{1998}).


\bibitem[{\citenamefont{Gulbahce et~al.}(2004)}]{Gulbahce2004}
\bibinfo{author}{\bibfnamefont{N.} \bibnamefont{Gulbahce}},
\bibinfo{author}{\bibfnamefont{H.} \bibnamefont{Gould}} \bibnamefont{and}
  \bibinfo{author}{\bibfnamefont{W.} \bibnamefont{Klein}},
  \bibinfo{journal}{Phys. Rev. E} \textbf{\bibinfo{volume}{69}},
  \bibinfo{pages}{036119} (\bibinfo{year}{2004}).

\bibitem[{\citenamefont{Zheng}(1998)}]{review:zheng98}
\bibinfo{author}{\bibfnamefont{B.}~\bibnamefont{Zheng}}, \bibinfo{journal}{Int.
  J. Mod. Phys. B} \textbf{\bibinfo{volume}{12}}, \bibinfo{pages}{1419}
  (\bibinfo{year}{1998}).

\bibitem[{\citenamefont{Gunton and Yalabik}(1978)}]{spinodal:gunton78}
\bibinfo{author}{\bibfnamefont{J.~D.} \bibnamefont{Gunton}} \bibnamefont{and}
  \bibinfo{author}{\bibfnamefont{M.~C.} \bibnamefont{Yalabik}},
  \bibinfo{journal}{Phys. Rev. B} \textbf{\bibinfo{volume}{18}},
  \bibinfo{pages}{6199} (\bibinfo{year}{1978}).

\bibitem[{\citenamefont{Mittag and Stephen}(1974)}]{potts:mittag74}
\bibinfo{author}{\bibfnamefont{L.}~\bibnamefont{Mittag}} \bibnamefont{and}
  \bibinfo{author}{\bibfnamefont{M.~J.} \bibnamefont{Stephen}},
  \bibinfo{journal}{J. Phys. A} \textbf{\bibinfo{volume}{7}},
  \bibinfo{pages}{L109} (\bibinfo{year}{1974}).

\bibitem[{\citenamefont{Klein}(1981)}]{spinodal:klein81}
\bibinfo{author}{\bibfnamefont{W.}~\bibnamefont{Klein}},
  \bibinfo{journal}{Phys. Rev. Lett.} \textbf{\bibinfo{volume}{47}},
  \bibinfo{pages}{1569} (\bibinfo{year}{1981}).

\bibitem[{\citenamefont{Kihara et~al.}(1954)\citenamefont{Kihara, Midzuno, and
  Shizume}}]{potts:kihara54}
\bibinfo{author}{\bibfnamefont{T.}~\bibnamefont{Kihara}},
  \bibinfo{author}{\bibfnamefont{Y.}~\bibnamefont{Midzuno}}, \bibnamefont{and}
  \bibinfo{author}{\bibfnamefont{T.}~\bibnamefont{Shizume}},
  \bibinfo{journal}{J. Phys. Soc. Japan.} \textbf{\bibinfo{volume}{9}},
  \bibinfo{pages}{681} (\bibinfo{year}{1954}).

\bibitem[{\citenamefont{Binder}(1981)}]{potts:binder81}
\bibinfo{author}{\bibfnamefont{K.}~\bibnamefont{Binder}}, \bibinfo{journal}{J.
  Stat. Phys.} \textbf{\bibinfo{volume}{24}}, \bibinfo{pages}{69}
  (\bibinfo{year}{1981}).

\bibitem[{\citenamefont{Meunier and Morel}(2000)}]{potts:meunier00}
\bibinfo{author}{\bibfnamefont{J.~L.} \bibnamefont{Meunier}} \bibnamefont{and}
  \bibinfo{author}{\bibfnamefont{A.}~\bibnamefont{Morel}},
  \bibinfo{journal}{Eur. Phys. J. B} \textbf{\bibinfo{volume}{13}},
  \bibinfo{pages}{341} (\bibinfo{year}{2000}).

\bibitem[{\citenamefont{Ferrero and Cannas}(2007)}]{potts:ferrero07}
\bibinfo{author}{\bibfnamefont{E.~E.} \bibnamefont{Ferrero}} \bibnamefont{and}
  \bibinfo{author}{\bibfnamefont{S.~A.} \bibnamefont{Cannas}},
  \bibinfo{journal}{Phys. Rev. E} \textbf{\bibinfo{volume}{76}},
  \bibinfo{pages}{031108} (\bibinfo{year}{2007}).

\bibitem[{\citenamefont{Peruggi et~al.}(1983)\citenamefont{Peruggi, di~Liberto,
  and Monroy}}]{potts:peruggi83}
\bibinfo{author}{\bibfnamefont{F.}~\bibnamefont{Peruggi}},
  \bibinfo{author}{\bibfnamefont{F.}~\bibnamefont{di~Liberto}},
  \bibnamefont{and} \bibinfo{author}{\bibfnamefont{G.}~\bibnamefont{Monroy}},
  \bibinfo{journal}{J. Phys. A} \textbf{\bibinfo{volume}{16}},
  \bibinfo{pages}{811} (\bibinfo{year}{1983}).

\bibitem[{\citenamefont{Wagner et~al.}(2000)\citenamefont{Wagner, Gresing, and
  Heide}}]{potts:wagner00}
\bibinfo{author}{\bibfnamefont{F.}~\bibnamefont{Wagner}},
  \bibinfo{author}{\bibfnamefont{D.}~\bibnamefont{Gresing}}, \bibnamefont{and}
  \bibinfo{author}{\bibfnamefont{J.}~\bibnamefont{Heide}}, \bibinfo{journal}{J.
  Phys. A} \textbf{\bibinfo{volume}{33}}, \bibinfo{pages}{929}
  (\bibinfo{year}{2000}).

\bibitem[{\citenamefont{Pearce et~al.}(1980)\citenamefont{Pearce and Griffiths}}]{pearce80}
\bibinfo{author}{\bibfnamefont{P.}~\bibnamefont{Pearce}},
\bibnamefont{and}
  \bibinfo{author}{\bibfnamefont{R.}~\bibnamefont{Griffiths}},
  \bibinfo{journal}{J.
  Phys. A} \textbf{\bibinfo{volume}{13}}, \bibinfo{pages}{2143}
  (\bibinfo{year}{1980}).

\end{thebibliography}

\newpage

\newpage

\newpage

\newpage

\newpage

\newpage

\newpage

\newpage

\newpage

\newpage

\newpage

\end{document}